# Communautés web 2.0 d'apprenants de langue avec parcours d'apprentissage : rôles, pédagogie et rapports au contenu

**Mathieu Loiseau\*, Anthippi Potolia\*\*, Katerina Zourou\*\*\***

*\* LIDILEM*
*Université Stendhal**, *BP 25*
*38040 Grenoble Cedex 9*
*mathieu.loiseau@msh-alpes.fr*
*\*\* PLIDAM*
*Inalco*
*49bis avenue de la Belle Gabrielle - 75012 Paris*
*apotolia@hotmail.com*
*\*\*\*Université du Luxembourg*
*FLSHASE/LCMI/DICA-lab*
*Route de Diekirch*
*7201 Walfer, Luxembourg*
*katerina.zourou@uni.lu*

*RÉSUMÉ. Parmi les nombreux types de collectifs d'apprenants de langues en ligne, cette contribution explore un type particulier de collectif, les communautés web 2.0 dotées de parcours d'apprentissage. Notre analyse se penche sur un échantillon composé de trois grandes communautés en termes d'effectifs (nombre d'utilisateurs et langues apprises): Babbel, Busuu et Livemocha. Elles sont examinées sous trois angles (rôles des utilisateurs, parcours pédagogiques et rapports au contenu) en vue de déterminer la structure technico-pédagogique et les affordances communicatives de ces espaces d'apprentissage collaboratif. Les tensions qui surgissent au niveau de la dimension technique (artefact web 2.0) et pédagogique (approches, démarches et ressources pédagogiques, rôles tuteurs-apprenants) seront abordées dans la partie finale.*

MOTS-CLÉS : apprentissage des langues, technologies émergentes, web 2.0, médias sociaux, télécollaboration.

## 1. Introduction

Dans le domaine de l'apprentissage des langues assisté par ordinateur (ALAO), de nombreux types de collectifs d'apprenants de langues en ligne, bénéficiant des possibilités d'interaction offertes par Internet, ont vu le jour depuis les années quatre-vingt : tele-tandem, e-twinning, plusieurs types de projets collaboratifs entre classes d'apprenants (et/ou de tuteurs). Ces projets suivent des modalités d'interaction et de collaboration variées, s'appuyant sur des artefacts numériques divers (et leur combinaison). Ces différentes initiatives ont finalement pu être observées et analysées sous des angles divers : modalités de télécollaboration, outils, médiation(s), démarches, aspects culturels, instrumentaux et sémiotiques, conditions de généralisation, etc. La littérature en ALAO est, notamment quant à ses paradigmes[1], abondante[2]. Toutefois, à l'heure actuelle peu de travaux ont, à notre connaissance, défriché les spécificités des mutations engendrées par le paradigme « web 2.0 » dans le champ de l'ALAO.

Les technologies du web 2.0 et surtout leur potentiel collaboratif (en termes de facilité de co-élaboration et de co-interprétation, de gestion et de valorisation de données et de ressources qui émanent d'un effort de partage et de co-construction) peuvent ouvrir de nouvelles pistes dans l'apprentissage des langues, que ce soit par le développement d'outils, mais aussi de nouveaux types de tâches, la mise en place de modes d'interaction novateurs ou simplement en utilisant le « buzz » autour du terme pour motiver / attirer les apprenants.

[CONOLE & ALEVIZOU 10:47-53] isolent neuf catégories d'activités du web 2.0 en recombinant celles proposées dans [CROOKS et al. 08:9-15]. Chaque activité y est associée à un outil et un seul. Il s'agit donc de l'activité centrale de chaque outil. Ils indiquent par exemple que les sites de réseaux sociaux ayant le plus de succès sont ceux qui permettent la création d'espaces dans lesquels les utilisateurs peuvent inviter des « amis », partager messages, textes, vidéos, annoter les ressources (tags) et jouer à des jeux (ce qui correspond pourtant à d'autres activités) [CONOLE & ALEVIZOU 10:49].

Dans le cadre de cette communication nous nous focalisons sur les outils de réseautage social (*social networking sites*) qui regroupent des sites comme *Facebook* ou *Myspace* ainsi que les outils de création d'espaces collectifs (entre autres : *Ning*, *Elgg*). [CACHIA 08:3] identifie six caractéristiques des sites de réseautage social correspondant largement à celles des sites communautaires que nous analysons : la valorisation du profil utilisateur (point d'entrée de ces sites, par lequel l'utilisateur se présente à la communauté) ; l'externalisation des données (Cachia parle avant tout des réseaux de contacts, mais celle-ci touche également les données propres aux objectifs du site, comme l'historique des contributions) ; l'émergence de nouveaux modes de constitution de réseaux (par exemple, last.fm met en contact ses utilisateurs selon les similitudes de leurs goûts musicaux) ; des activités « bottom-up » (les utilisateurs influencent le contenu, cf. 3.3). Elle conclut son énumération par deux critères que nous n'aborderons pas : facilité d'utilisation et réorganisation de la géographie d'Internet. Nous montrerons à travers notre description des différentes communautés analysées qu'elles s'inscrivent dans le paysage des technologies du web 2.0.

---

[1] Pour un aperçu voir aussi [ZOUROU 07].

[2] Cf. [CHAPELLE97] [LAMY & HAMPEL 07] [LUND 03] [MANGENOT 02] pour n'en mentionner que quelques unes.

*1.1. Délimitation du terrain d'analyse*

Deux critères ont été retenus dans la délimitation du terrain d'analyse. Premièrement, les autres types de technologies web 2.0 (ex : mondes virtuels) sont exclus : ils n'offrent pas les mêmes caractéristiques fonctionnelles que les sites de réseau social (ils en sont d'ailleurs séparés dans [CONOLE & ALEVIZOU 10] et [CROOKS et al. 08]) et vu sous le prisme de l'instrumentation [RABARDEL 95], leur potentiel interactif n'est pas comparable.

Ensuite, seules les communautés affichant un objectif explicite d'apprentissage des langues suivant une progression explicite plus ou moins élaborée ont été retenues (même si, comme nous le verrons par la suite, cet objectif est revisité par les tensions entre les différents acteurs). Seront donc exclues les communautés Facebook sans aide à l'apprentissage (ex : communautés Facebook de la BBC, de RFI). Même si l'un des objectifs de ces collectifs est la pratique de la langue, elle n'est pas inscrite dans une démarche pédagogique intégrant les étapes d'une progression. Si dans certains cas, des activités peuvent être proposées par des enseignants pour leurs apprenants, les ressources pédagogiques et la modélisation sous-jacentes ne sont accessibles qu'à un sous-groupe d'utilisateurs constitué préalablement.

*1.2. Démarche méthodologique*

La démarche méthodologique est fondée sur la conception de trois critères d'analyse : rôle des utilisateurs, modélisation pédagogique et horizontalité. Ils sont proposés comme point de départ pour établir une batterie d'outils méthodologiques amenée à évoluer avec l'analyse des différentes plateformes et les pratiques elles-mêmes. S'agissant d'une démarche exploratoire, les critères proposés ne sont pas exhaustifs et doivent être compris comme une tentative de proposer des pistes méthodologiques aptes à saisir l'originalité de ces communautés et à souligner le processus analytique découlant de cette originalité : leur appartenance au paradigme « web 2.0 ».

**2. Communautés web 2.0 d'apprentissage des langues : bref état des lieux**

Une première tentative de recenser les communautés web 2.0 d'apprentissage des langues et d'en identifier les spécificités interactionnelles a été réalisée en amont, au sein d'une étude plus large [DIXHOORN et al. 10]. Ce travail préliminaire visant à identifier la manière dont les concepteurs de ressources se servent du web 2.0 pour faciliter et renforcer l'apprentissage des langues ou leur position comme acteur du domaine, nous a permis d'effectuer un premier défrichage du potentiel pédagogique de ces collectifs, malgré le fait que le web 2.0 en tant que construction conceptuelle-idéologique et même technique (cf. évolution rapide) soit loin d'être défini et partagé unanimement (voir [DILGER 10] et pour l'éducation [ANDERSON 07] [GOODFELLOW & LAMY 10]).

Parmi les communautés web 2.0 d'apprentissage des langues, nous nous penchons – rappelons-le – sur celles contenant des parcours pédagogiques, en excluant les deux autres types de communautés (sans parcours pédagogique et de type *marketplace* [LOISEAU et al. 10]). Cette précision nous amène à la définition de notre terrain d'analyse, composé de trois communautés web 2.0 d'apprentissage des langues proposant des parcours d'apprentissage parmi les plus importantes : *Babbel*, *Busuu* et *Livemocha*[3]. Ce choix est conditionné par leur impact en termes de nombre de membres et de langues apprises[4]. Pour chaque site nous avons analysé avant tout l'offre qui nous paraissait la plus représentative de la communauté. *Babbel* étant avant tout un site payant, nous nous sommes abonnés pour pouvoir le tester. *Busuu* et *Livemocha* sont plutôt pourvus d'un modèle économique de la gratuité mêlant formule *freemium* et financement publicitaire[5]. Dans ces modèles, la rentabilité vient avant tout de l'offre gratuite : le *freemium* nécessite une offre basique accessible à tous menant certains utilisateurs à souscrire à une offre payante plus complète [DILGER 10], alors que le modèle publicitaire dépend évidemment de la fréquentation du site [WAUTHY 08]. Nous nous sommes focalisés sur les offres gratuites de ces deux acteurs.

Avant de lancer une étude à plus grande échelle, il convient de tester nos critères sur un petit nombre de communautés non seulement pour voir s'ils sont fonctionnels, mais également en vue de les prolonger du fait des questions qu'ils peuvent soulever.

### 2.1. Fonctionnement archétypique

Avant de passer aux critères eux-mêmes, il convient d'expliquer le fonctionnement global des sites, dont l'un des dénominateurs communs est de permettre à l'utilisateur de jouer tantôt le rôle d'apprenant, tantôt celui de tuteur. Chaque communauté a sa propre conception de parcours, son acception de la didactique et de la pédagogie, qui proviennent avant tout du cadre dans lequel elle évolue. Les modalités d'interaction et le type d'activité varient d'une plateforme à l'autre mais on retrouve les briques élémentaires suivantes : – la leçon (ou présentation des notions à acquérir) ; – des exercices structuraux ou calqués sur le modèle de la leçon, qui sont corrigés automatiquement ; – des exercices de production écrite et orale, corrigés par les utilisateurs (ceux qui acceptent d'endosser le rôle de tuteur) ; – des modalités d'interaction synchrones ou asynchrones avec d'autres utilisateurs (tuteurs) dans le cadre de la séquence en utilisant les outils de communication proposés par la plateforme.

Pour [MUSSER et al. 07:10] l'effet de réseau est une propriété majeure du web 2.0[6]. Or, il s'exprime principalement dans ces communautés via la possibilité pour les apprenants d'obtenir une (et parfois plusieurs) correction personnalisée de chaque exercice de production, ce qui représente l'un des intérêts majeurs de ce type de plateforme. Ces corrections sont faites par des utilisateurs qui ont la plupart du temps rejoint la plateforme pour **apprendre** les langues. La question qui se pose alors est de savoir comment ils sont encouragés à endosser le rôle de tuteur. Cette question nous mène tout droit à notre premier critère, à savoir l'analyse du rôle des utilisateurs dans les trois communautés identifiées.

---

[3] Respectivement : http://fr.babbel.com/, http://www.busuu.com/fr et http://fr-fr.livemocha.com.

[4] Chiffres recueillis en janvier 2011. *Babbel* : 1 million de membres sur son site, 4.000 membres Facebook et sept langues proposées (EN, FR, DE, IT, ES, PT, SV). *Busuu* : 2 millions de membres, 31.500 membres Facebook (depuis son lancement, 09/2010), 7 langues offertes (EN, FR, DE, IT, ES, PT, RU). *Livemocha* : 8,5 millions de membres (de 200 pays), 163.000 membres Facebook et offre de 35 langues (dont des langues (très) rares y compris l'arabe, le catalan, l'esperanto, le grec, l'estonien…).

[5] Pour une description concise de chacun de ces modèles, voir http://bit.ly/Guillaud.

[6] Cf. leur définition : Le web 2.0 est « […] un ensemble de courants sociaux, économiques et technologiques qui forment collectivement la base de la prochaine génération de l'Internet : un media plus mature, à part, dont les caractéristiques sont : participation des utilisateurs, ouverture et effets de réseau » (notre traduction).

## 3. Analyse des communautés

Après avoir analysé le rôle des utilisateurs pour chacune des communautés de notre étude, nous examinerons deux autres critères. L'un est tourné vers la dimension pédagogique sous-jacente aux parcours dont nous détaillerons la modélisation, alors que l'autre tient plus à une autre notion centrale du web 2.0, le contenu généré par les utilisateurs (CGU). Pour chacun de ces critères et pour s'assurer qu'ils fonctionnent, nous allons tenter de placer chacune des communautés étudiées sur un continuum.

### *3.1. Premier critère : rôle des utilisateurs*

Le premier critère d'analyse que nous avons retenu est celui du statut des utilisateurs, à savoir leur obligation ou non d'endosser un rôle de tuteur.

*Babbel*, qui pose le moins de contraintes à ce sujet, n'oblige aucunement ses utilisateurs à assumer la fonction de tuteur. *Babbel* propose certes à ses membres de se répartir en sous-communautés de co-apprenants (organisées par couples L1-L2) et de partenaires de tandem (fonction classique d'un apprentissage en tandem [LEWIS & STICKLER 07]) dans le but de mettre en pratique leurs acquis, s'entraider, trouver des solutions à leurs problèmes et socialiser leur apprentissage. Cependant, la participation à ces sous-communautés est laissée à la discrétion des utilisateurs : elle n'est pas promue et à l'inverse la non-participation n'est pas sanctionnée. Le seul critère pris en compte pour évaluer l'activité des utilisateurs est leur parcours pédagogique, c'est-à-dire le travail utilisant les contenus de *Babbel*. L'esprit d'entraide, la correction des productions des autres apprenants, la disponibilité, bref l'esprit de communauté sont quant à eux laissés à la discrétion des utilisateurs de *Babbel*.

Dans le cas de *Busuu* la situation est sensiblement différente : ce sont les services rendus à la communauté (et non l'activité d'apprentissage) qui, sans être obligatoires, sont valorisés et récompensés par un système de points *berries* : plus on offre ses services à la communauté, plus on récolte de *berries* affichées d'office sur son profil. La conséquence implicite de ce fonctionnement est que les utilisateurs les plus « importants » de la communauté, ou perçus comme tels, sont ceux dont le profil exhibe le plus de *berries*. Parmi les services ouvrant le droit à des *berries* figurent : commenter des exercices d'écriture, se constituer un réseau d'amis, créer un groupe thématique et modérer un groupe actif, traduire des pages de *Busuu.com*, proposer des idées pour améliorer le site.

*Livemocha* se rapproche de *Busuu* : l'utilisateur récolte des *mochapoints* en fonction de son activité d'apprenant et à chaque fois qu'il prête main-forte à la communauté. *Livemocha* tient également un décompte d'un « score d'enseignant » (portion des *mochapoints* glanés par l'utilisateur en endossant le rôle de tuteur). Ce type de points a la particularité d'ouvrir des droits, comme l'accès à tous les cours de toutes les langues (payants autrement) ou de devenir traducteur du site (un autre moyen de gagner des *mochapoints*). Par ailleurs, l'invitation à l'entraide et à la responsabilisation communautaire, notamment par la correction des productions des autres membres, est plus explicite chez *Livemocha* :

> « En classant et en corrigeant d'autres soumissions de membres de la communauté vous <u>garantissez un environnement d'apprentissage productif pour vous et les autres membres de la communauté</u>. Voici ci-dessous des soumissions de membres de la communauté <u>essayant d'apprendre</u> <u>votre</u> langue ».
> http://fr-fr.Livemocha.com/tutor/language/index

Environnements Informatiques pour l'Apprentissage Humain, Mons 2011

Comme nous l'avons indiqué, la possibilité pour l'utilisateur de voir l'essentiel de ses productions commentées est l'une des composantes centrales des communautés d'apprenants. Cependant, cette spécificité est exploitée différemment par les concepteurs des communautés retenues. *Livemocha* est celle qui cherche à tirer profit le plus explicitement de cette caractéristique pour dynamiser l'effet de réseau et construire – d'après Clint Schmidt, vice-président marketing de *Livemocha* quand il nous a accordé un entretien[7] – des « cercles vertueux » : l'utilisation des points (comme dans *Busuu* mais avec la formule paiement ou participation) pour qualifier les utilisateurs doit influencer la tâche de correction des autres utilisateurs. Comme il s'agit d'une tâche fastidieuse, ils ne vont pas nécessairement corriger un grand nombre de productions. Les concepteurs anticipent que ces corrections concerneront en particulier le réseau de contacts du tuteur et les utilisateurs les plus « méritants ». En d'autres termes, un utilisateur qui ne corrige jamais les productions des autres, porte cette inactivité dans son profil et risque de s'isoler à terme, les autres membres – correcteurs potentiels – choisissant de privilégier alors la correction des productions à leurs contacts et ceux qui participent activement à la vie de la communauté.

Pour revenir aux critères de [CACHIA 08], cette première analyse montre l'utilisation du profil comme interface de présentation à la communauté (points). Elle permet également d'aborder les modes de constitution d'un réseau propres à ce type de communauté, particulièrement évidents dans *Livemocha*, où pour encourager ce « cercle vertueux », l'apprenant se voit proposer de changer de rôle après chaque activité. Il a aussi la possibilité de solliciter des utilisateurs, en priorité parmi ses contacts pour corriger sa production. D'après notre expérience, ceux-ci semblent être dans l'extrême majorité des cas des utilisateurs « rencontrés » en corrigeant leurs productions ou suite à leur correction de la nôtre.

Enfin, le « mérite » des enseignants de *Livemocha* n'est pas que quantitatif, puisque chaque retour d'enseignant est évalué qualitativement par les membres de la communauté (notes). Ceci met en exergue un problème apparent : l'apprentissage n'est encadré que par des apprenants endossant le rôle de tuteur. Or la qualité des retours est nécessairement un problème que les communautés devront se poser, que ce soit pour se satisfaire de la formule actuelle ou entreprendre une forme de professionnalisation de la partie enseignement.

*3.2. Deuxième critère : degré de modélisation pédagogique*

Les utilisateurs ne sont pas les seuls tenants de la modélisation pédagogique en vigueur dans la communauté : les supports utilisés dans l'apprentissage émanent des concepteurs des sites. Il s'agit, pour le présent critère, d'aller au-delà du strict *modus operandi* et nous intéresser à la manière dont les différents acteurs conçoivent l'accès au savoir linguistique. La modélisation pédagogique sera examinée sous l'angle de la cohérence des parcours proposés et des principes régissant ces derniers en termes de choix didactiques.

Dans le cas de *Babbel*, il n'y a pas à première vue de progression au sens habituel du terme. La structuration des contenus se fait suivant différents découpages grammaticaux, situationnels, lexicaux ou bien par compétences et par besoins (utilité, rapidité, répétition) :

---

[7] [DIXHOORN et al. 10] visait notamment à recueillir le point de vue des concepteurs de ressources sur les mutations accompagnant l'avènement de l'expression « web 2.0 ». Une partie des informations recueillies s'est faite par entretiens semi-dirigés.

- Thèmes et situations
- Cours express
- Exercices de répétition
- Écouter et pratiquer
- Exercices d'écriture
- Grammaire
- Vocabulaire de base et avancé
- 1000 phrases utiles

Guidé uniquement par les ressources disponibles et leur organisation, l'utilisateur est maitre de son apprentissage. On part donc du principe que l'utilisateur a la distance et la compétence nécessaires pour définir ses besoins, évaluer ses lacunes et construire son propre parcours d'apprentissage. Cette représentation de l'utilisateur change dans le cadre des cours pour débutants. En effet, *Babbel* opte dans ce cas pour une structuration fonctionnelle (par actes de parole) des contenus, similaire aux manuels de langue papier :

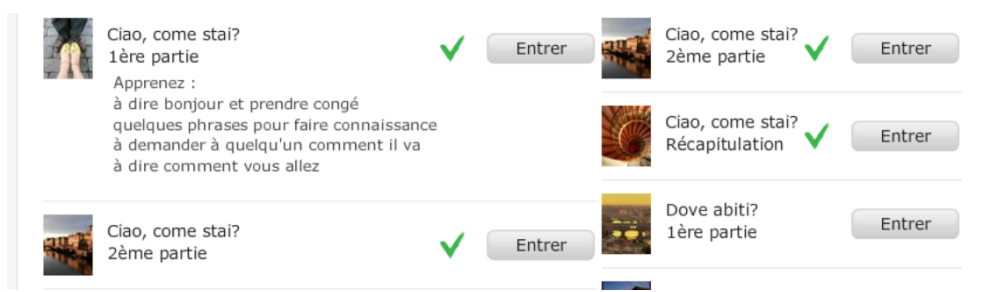

**Figure 1. Structuration fonctionnelle des ressources dans Babbel**[8]

Un autre élément en matière de structuration de contenus – trace de l'adaptation progressive des instances conceptrices des communautés aux besoins réels ou envisagés de leurs membres – mérite d'être signalé. Depuis décembre 2010, *Babbel* propose une nouvelle répartition par trimestre de ses modules pédagogiques. Il ne s'agit pas de nouveaux contenus mais d'une redistribution des anciens sur une année (en modules trimestriels) faite dans un souci d'accompagnement de l'utilisateur à travers les étapes de son apprentissage :

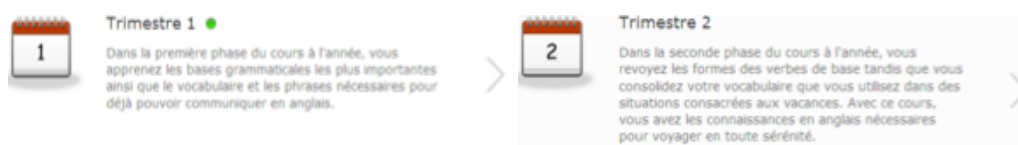

**Figure 2. Répartition des modules par trimestre dans Babbel**

*Babbel* viserait à travers la structuration d'ensemble de ses contenus un public avec des profils et des besoins différenciés en accord *a priori* avec les principes régissant les situations d'auto-apprentissage. Cependant, les unités, leçons ou cours proposés sont en fait des *phrasebooks* : la leçon est un ensemble de phrases (ou mots) regroupés selon un critère (thème, grammaire, etc.). La répétition de phrases ou mots hors contexte, leur traduction en L1, leur orthographe y supplantent les tendances actuelles en didactique des langues.

Il en va de même pour *Livemocha* où le principe des *phrasebooks* associés à des points grammaticaux ou à des thèmes est le seul mobilisé. Ils semblent viser à terme l'acquisition de certaines compétences (ex : cours d'anglais débutants (unité 104), il est annoncé qu'au terme de l'unité les étudiants seront « capables de se présenter et de comprendre certaines instructions ») mais rien dans la méthodologie mise en place ne permet de les atteindre. Le choix des *phrasebooks* comme unique source d'apprentissage « interdit » l'usage de documents authentiques et même de textes ou de dialogues comme documents déclencheurs de compréhension, de pratique grammaticale, de contextualisation des dimensions fonctionnelle et notionnelle de la langue étudiée. Les exercices structuraux prennent alors le dessus reflétant un apprentissage behavioriste d'une langue étrangère.

---

[8] Les figures recombinent parfois plusieurs portions de saisie d'écran, afin de diminuer l'espace mobilisé.

On peut également noter qu'un écart important se creuse entre les parties compréhension (*phrasebook*) et production. Si en termes d'*input* on se limite à une exposition à des mots ou des phrases, en termes d'*output*, des productions (écrites) plus complexes sur le plan discursivo-textuel sont attendues (ex : Décrivez la météo pour cinq jours et expliquez ce dont différentes personnes auront besoin selon le temps.[9]). C'est aussi le cas de *Babbel*.

Cependant le défaut de modélisation de *Livemocha* le plus grave à notre sens provient de la volonté de proposer un nombre important de langues à apprendre (35). La progression pour ces langues est calquée sur les spécificités morphologiques de l'anglais. Les contenus pédagogiques des 34 autres langues sont alors de simples traductions des leçons prévues pour l'anglais. Cela a de lourdes conséquences (ex : comme en anglais les mots n'ont pas de genre, en russe on apprend à dire « elle est vieille » – она старая – mais pas « il est vieux » – он старый). Ainsi le « remixage » des données prôné par [MUSSER et al. 07:24] ne peut être appliqué aveuglément, au risque de l'être en dépit du bon sens pédagogique.

D'un point de vue macrostructural l'organisation et la répartition *a priori* raisonnées des contenus pédagogiques en cours, unités, et enfin leçons souffrent de certains défauts. En effet, les cours sont désignés par un code peu transparent (ex : « Anglais 102 », « Anglais 202 », etc.). Même si ce code est suivi d'un court texte qui cherche à expliciter le public visé, il nous semble que cet objectif louable n'est pas atteint pour autant.

Quant aux leçons, celles-ci ne sont pas construites sur des objectifs formulés de manière identique et alternent donc entre visées grammaticales (« Unité 9, Leçon 2 : Verbes réguliers du présent progressif/passé simple »), lexicales («Unité 4, Leçon 1 : Nombres ») ou bien lexico-communicatives (« Unité 4, Leçon 5 : Directions »).

*Busuu* est le seul des trois sites examinés à faire référence au *Cadre Européen Commun de Référence* (*CECR*) et à organiser ses contenus suivant celui-ci (de A1 à B2).

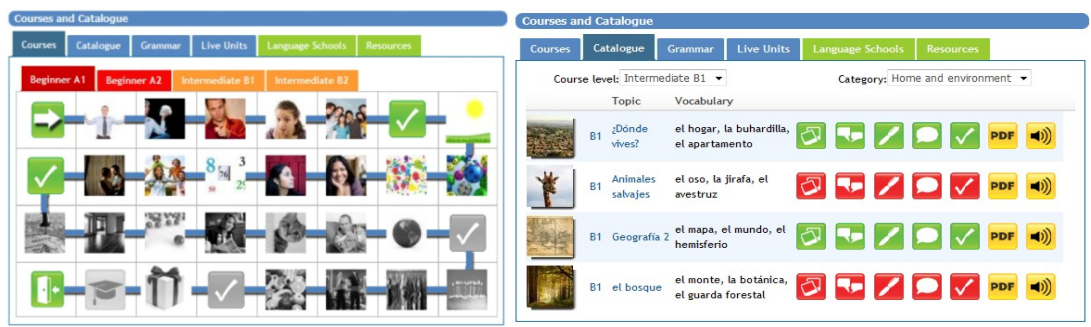

**Figure 3. Deux représentations différentes de la progression dans Busuu**

Pour chaque niveau, l'utilisateur peut choisir un accès au savoir par cours, par catalogue ou par thème grammatical. La progression par « cours » est la plus linéaire : un apprentissage « à la carte », similaire aux cours par trimestre de *Babbel*. L'utilisateur peut visualiser sous forme de chemin toutes les étapes à accomplir et franchir pour atteindre le niveau choisi (cf. Figure ). Le parcours par « catalogue », plus libre, repose sur le principe que l'utilisateur est à même d'identifier ses lacunes, comme ses besoins, et de construire lui-même sa progression à partir d'indications de niveau, de vocabulaire et de thème (cf. Figure ). Enfin, le parcours par « thème grammatical » fait partie de la version *Premium* de *Busuu*.

---

[9] http://fr-fr.*Livemocha*.com/exercises/view/447

Par ailleurs, *Busuu* est le seul site qui s'écarte des *phrasebooks*. On reste dans un univers relativement aseptisé culturellement et inférieur à certains matériels d'apprentissage d'une langue étrangère diffusés sur supports « classiques ». Cependant, contrairement à *Babbel* et à *Livemocha*, *Busuu* réintroduit dans l'apprentissage le dialogue en tant que support de compréhension et qu'*input* plus complexe et pluriel que la formule *phrasebooks*, qui existe toujours ici mais qui devient partie d'une étape située en amont du dialogue remplissant alors une fonction préparation à la réception et appropriation des contenus du dialogue.

Au terme de l'étude du critère de modélisation pédagogique, dont il convient de rappeler qu'elle porte sur les supports et non sur l'activité globale de l'apprenant dans le cadre de ces sites, il nous semble que *Busuu* propose le contenu le plus en accord avec la didactique des langues actuelle. L'environnement pédagogique de *Busuu* n'est pas pour autant idéal en termes de cohérence pédagogique mais il reste le moins perméable à la représentation de l'apprentissage d'une langue comme une liste de mots et de structures à mémoriser.

### *3.3. Troisième critère : horizontalité*

Dans ce dernier critère testé, la dimension des usages, au moins tels qu'ils sont prescrits, est abordée sous le prisme d'une des caractéristiques centrales de la notion de web 2.0 que l'on pourrait appeler l'horizontalité. Elle a trait aux activités « bottom-up » évoquées par [CACCHIA, 08] ou au CGU, dont l'importance en tant que concept peut être attestée par sa position pivot dans la définition média social de [KAPLAN & HAENLEIN 10:61] : « un groupe d'applications Internet qui se fonde sur les fondations idéologiques et technologiques du web 2.0 et qui permet la création et l'échange de CGU »[10].

La première question à nous poser pour pouvoir parler de l'horizontalité du site est d'en définir le contenu. Nous n'analysons pas la gouvernance du site ou la possibilité pour un utilisateur d'influencer indirectement (même drastiquement) la plateforme (ex : suggestion de fonctionnalités), mais d'en affecter directement le contenu et ses différents types. Nous avons séparé les données du site en trois catégories (qui pourraient être placées selon un axe).

<u>Le contenu principal</u> est celui qui est au cœur de la plateforme et autour duquel gravitent les interactions d'apprentissage des utilisateurs. Nous considérerons ici qu'il s'agit du matériel pédagogique tel qu'il est intégré à la progression des utilisateurs. Dans les entretiens que nous avons menés (cf. [DIXHOORN et al. 10]), il s'agissait du sens que donnaient implicitement les concepteurs à ce mot.

Parmi les sites communautaires analysés, seul *Livemocha* permet aux utilisateurs d'avoir un impact direct sur le contenu. Cependant, les utilisateurs ne peuvent pas réellement en ajouter : le contenu qu'ils fournissent n'est que la traduction des supports pour l'anglais (avec les conséquences pointées dans la section 3.2). En effet, les utilisateurs ayant acquis un statut les y autorisant peuvent proposer des traductions des interfaces du site ou des ressources d'apprentissage pour leur L1. Dans le dernier cas, il s'agira de contenu à proprement parler, mais pas dans le premier. Un certain nombre de cours pour des langues plus ou moins « rares » ont vu le jour de la sorte (cf. section 3.2). Les contributeurs y sont mentionnés. Cependant on ne peut pas parler de CGU au même sens que dans la *Wikipédia* par exemple, où il n'est pas prédéterminé, l'utilisateur décidant des contenus informationnel et structurel des articles auxquels il contribue.

---

[10] Notre traduction.

Le contenu « annexe » ne l'est pas nécessairement du point de vue de l'importance qu'il revêt pour les différents utilisateurs, il est cependant annexe dans le sens où il n'est pas intégré au contenu principal, à la progression. C'est par exemple le cas des *flashcards Livemocha*[11], qui sont destinées à un usage « désintéressé » de l'apprenant, qui ne gagnera pas de *mochapoints* à s'entrainer avec celles-ci, alors qu'il en gagnera pour en créer. Mais les leçons ne renvoient pas à ces fiches. De la même manière les groupes *Busuu* présentent des données persistantes notamment sur des problèmes culturels, mais il n'y a pas plus d'intégration de ces données aux séquences d'enseignement et très peu de possibilités de structuration du contenu. Il s'agit d'un usage de forums traditionnels comme source de données plus organisée (et donc plus apte à être réutilisée) que le « tableau » de *Babbel*, qui conserve des échanges culturels, linguistiques ou pratiques des utilisateurs.

Les traces, enfin, seront des données persistantes qui peuvent être accessibles à tous, comme les corrections d'exercices de production, ou seulement à leurs auteurs, comme les réponses aux exercices de *Busuu* qui sont conservées et peuvent servir de source pour la génération automatique d'exercices (l'utilisateur peut spécifier si l'exercice doit être composé uniquement de « ses erreurs », de « son vocabulaire marqué » ou s'il doit s'agir d'une « révision aléatoire »). Il ne s'agit pas de contenu dans la mesure où les traces ne sont pas conçues pour être une source d'information pour la communauté (ce qui serait différent par exemple avec un système de notation des échanges pour la correction d'un exercice, en fonction duquel pourraient être proposés comme exemples de correction accessibles directement depuis celui-ci). N'ayant accès aux mêmes données que les concepteurs, nous devons nous contenter d'une acception générique du terme, et plus proche du sens étymologique que de celui de l'ingénierie des traces [LAFLAQUIÈRE et al., 2007]. Notre définition satisfait cependant notre statut d'observateur et permet de considérer aussi bien les « marques laissées par une action » (cf. corrections de productions), que les traces utilisées pour adapter l'environnement aux utilisateurs (exercices dans *Busuu*).

Enfin, nous ne considérons pas le cas des échanges synchrones (données non persistantes).

Suite à ces remarques, nous nous proposons de décrire l'horizontalité comme un continuum du degré d'exploitation des données générées par les utilisateurs dans le site, sur le modèle de la description de la propriété d'un site « d'être web 2.0 » de [BONDERUP DOHN 09]. N'ayant pas d'information sur les statistiques de fréquentation, nous fondons cette échelle sur les différents types de contenus que nous avons définis. Le CGU concerne à notre sens en premier lieu le contenu principal du site et même si du point de vue des usages le « contenu annexe » pourrait s'avérer plus utilisé que le contenu principal, il ne semble pas que ce soit le cas. De plus, du point de vue conceptuel, c'est ce dernier qui est au centre de la conception des sites et tant que la place du premier n'est pas réévaluée et ses données mieux intégrées et structurées, il ne peut, à notre sens, présider au degré d'horizontalité. Cette échelle nous permet de traiter via le même critère le contenu au sens le plus strict et le moins strict, prenant ainsi en compte la variété des utilisations.

---

[11] Exercice de révision lexical où l'apprenant se voit poser circulairement un ensemble de questions jusqu'à avoir répondu à toutes (ou l'écoulement d'un temps prédéfini).

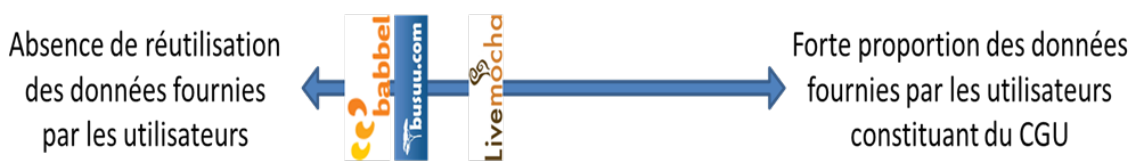

**Figure 4. Représentation des trois communautés sur un axe selon notre critère d'horizontalité**

Ainsi, *Livemocha*, bien que peu horizontal dans sa conception (on ne peut réellement parler de CGU) est la seule des communautés visitées où les contributions des utilisateurs peuvent toucher le contenu principal. Les trois communautés ont une forme de contenu annexe pédagogique ou utilisable pour l'apprentissage, nous ne pouvons donc les séparer réellement de ce point de vue, mais seul *Busuu* semble utiliser les traces pour proposer à l'utilisateur un contenu plus adapté, d'où sa position devant *Babbel* dans la Remarques/conclusion

Les trois critères proposés ici ont été traités indépendamment. Bien qu'ils nous semblent constituer des outils d'analyse pertinents, ils soulèvent de nombreuses questions.

Ils nous ont par exemple forcés à aborder les contenus à travers un filtre, permettant de dissocier le contenu principal du site du contenu annexe.

La qualité des corrections et autres contributions, ainsi que les systèmes d'évaluation par points posent la question de la professionnalisation des tuteurs. Dans ce cas précis la qualité de natif est-elle suffisante, en particulier avec un contenu principal dépourvu d'explication métalinguistique (cf. section 3.2) ? Le nombre peut-il compenser l'absence d'expertise ? Ou faut-il s'engager plus avant dans la certification ? *Livemocha*, qui fait déjà intervenir les notations dans l'accès aux fonctionnalités et dans les droits d'accès, lance un statut d'enseignant coopté par la plateforme, qui permettra aux enseignants les plus appréciés et les plus actifs (éventuellement sous conditions d'expérience à l'extérieur du site) d'accéder à des cours, voire d'être payés pour leurs interventions.

Enfin, c'est la conjonction de plusieurs critères qui fait peut-être naitre les questions les plus intéressantes du point de vue de la recherche. Nous avons constaté que dans le cas de *Livemocha*, la plus grande réutilisation des ressources se faisait au détriment de la modélisation pédagogique. On est cependant en droit de faire l'hypothèse que cette influence n'est pas systématique. On peut imaginer l'exemple d'un véritable CGU, qui permettrait aux utilisateurs de diversifier les énoncés d'exercices ou de restructurer les cours (lors de traductions, qui n'en seraient plus vraiment), et qui de ce fait pourrait, sous certaines conditions, profiter à la modélisation pédagogique du contenu.

Étudier ces critères en corrélation les uns avec les autres, par exemple en termes d'importance relative au moment de la conception ou en termes de l'influence qu'ils ont les uns sur les autres pourrait permettre de concevoir des outils d'analyses plus complets. Ceux-ci constitueraient des aides non seulement pour la catégorisation des plateformes, mais ils pourraient permettre de proposer des moyens d'utiliser des outils du web 2.0 pour l'enseignement des langues, ou encore des modèles de conception où les aspects techniques seraient au service d'une modélisation pédagogique pleinement consciente de leurs possibilités.

**Remerciements**

## 4. Bibliographie